\documentclass[letterpaper,11pt]{article}
\usepackage{graphicx}
\usepackage{lineno}
\input babarsym
\def\Kstar                   {\ensuremath{K^{*}}\xspace}
\def\Kstarp                  {\ensuremath{K^{*+}}\xspace}

\def\dz                      {\ensuremath{d_{\scriptscriptstyle 0}\xspace}}

\def\Cerenkov                {\v{C}erenkov}

\newcommand{\BABARPubYear}    {06}

\newcommand{\BABARConfNumber} {025}
\newcommand{\SLACPubNumber} {11979}

\long\def\inst#1{\par\nobreak\kern 4pt\nobreak
    {\it #1}\par\vskip 10pt plus 3pt minus 3pt}

\begin{document}
{\pagestyle{empty}

\begin{flushright}
\babar-CONF-\BABARPubYear/\BABARConfNumber \\
SLAC-PUB-\SLACPubNumber \\
July 2006 \\
\end{flushright}

\par\vskip 5cm

\begin{center}
\Large \bf Search for the Rare Decay \\
          \boldmath \Bpmtoazpmpiz\\ 
\end{center}
\bigskip

\begin{center}
\large The \babar\ Collaboration\\
\mbox{ }\\
\today
\end{center}
\bigskip \bigskip

\begin{center}
\large \bf Abstract
\end{center}
A search for the decay $B\pm \to a_{0}^{\pm} \pi^{0}$
with the $a^{+}_{0}$ decaying to an $\eta$ and a $\pi^{+}$ was carried out
at the Stanford Linear Accelerator Center using the {\mbox{\sl B\hspace{-0.4em} {\small\sl A}\hspace{-0.37em} \sl B\hspace{-0.4em} {\small\sl A\hspace{-0.02
em}R}}} detector
coupled with the PEP-II collider.  The analysis used a data sample
comprised of approximately 252 million $B {\kern 0.18em \overline{\kern -0.18em B}{}\xspace}$ pairs collected at the
$\Upsilon$(4S) resonance. No signal was observed and a 90$\%$
confidence level upper limit on the branching fraction was set at
$1.32 \times 10^{-6}$.

\vfill
\begin{center}

Submitted to the 33$^{\rm rd}$ International Conference on High-Energy Physics, ICHEP 06,\\
26 July---2 August 2006, Moscow, Russia.

\end{center}

\vspace{1.0cm}
\begin{center}
{\em Stanford Linear Accelerator Center, Stanford University, 
Stanford, CA 94309} \\ \vspace{0.1cm}\hrule\vspace{0.1cm}
Work supported in part by Department of Energy contract DE-AC03-76SF00515.
\end{center}

\newpage
}

\begin{center}
\small

The \babar\ Collaboration,
\bigskip

%
{B.~Aubert,}
{R.~Barate,}
{M.~Bona,}
{D.~Boutigny,}
{F.~Couderc,}
{Y.~Karyotakis,}
{J.~P.~Lees,}
{V.~Poireau,}
{V.~Tisserand,}
{A.~Zghiche}
\inst{Laboratoire de Physique des Particules, IN2P3/CNRS et Universit\'e de Savoie,
 F-74941 Annecy-Le-Vieux, France }
% ...

\end{center}\newpage

\section{INTRODUCTION}
\label{sec:introduction}
The quark content of the $a_{0}$ mesons is a subject of debate.  It
has been conjectured that they may not be simple $q \overline q$
states but may have a more complex nature~\cite{ref:baru2003} such as
having a $K\overline K$ component, being a glueball, or being a
mixture of two and four quark states. According to Delepine {\em et
al.}~\cite{ref:delepine2005}, a measurement of the branching fraction
(BF) of \Btoazpiz\ may help in resolving this problem, as the
predicted branching fraction for the four quark model is expected to
be up to an order of magnitude less than the two quark model. The
predicted branching fraction for this is already small, of order $2
\times 10^{-7}$.

\begin{figure}[htbp] 
\begin{center}
\includegraphics[width=6.5cm]{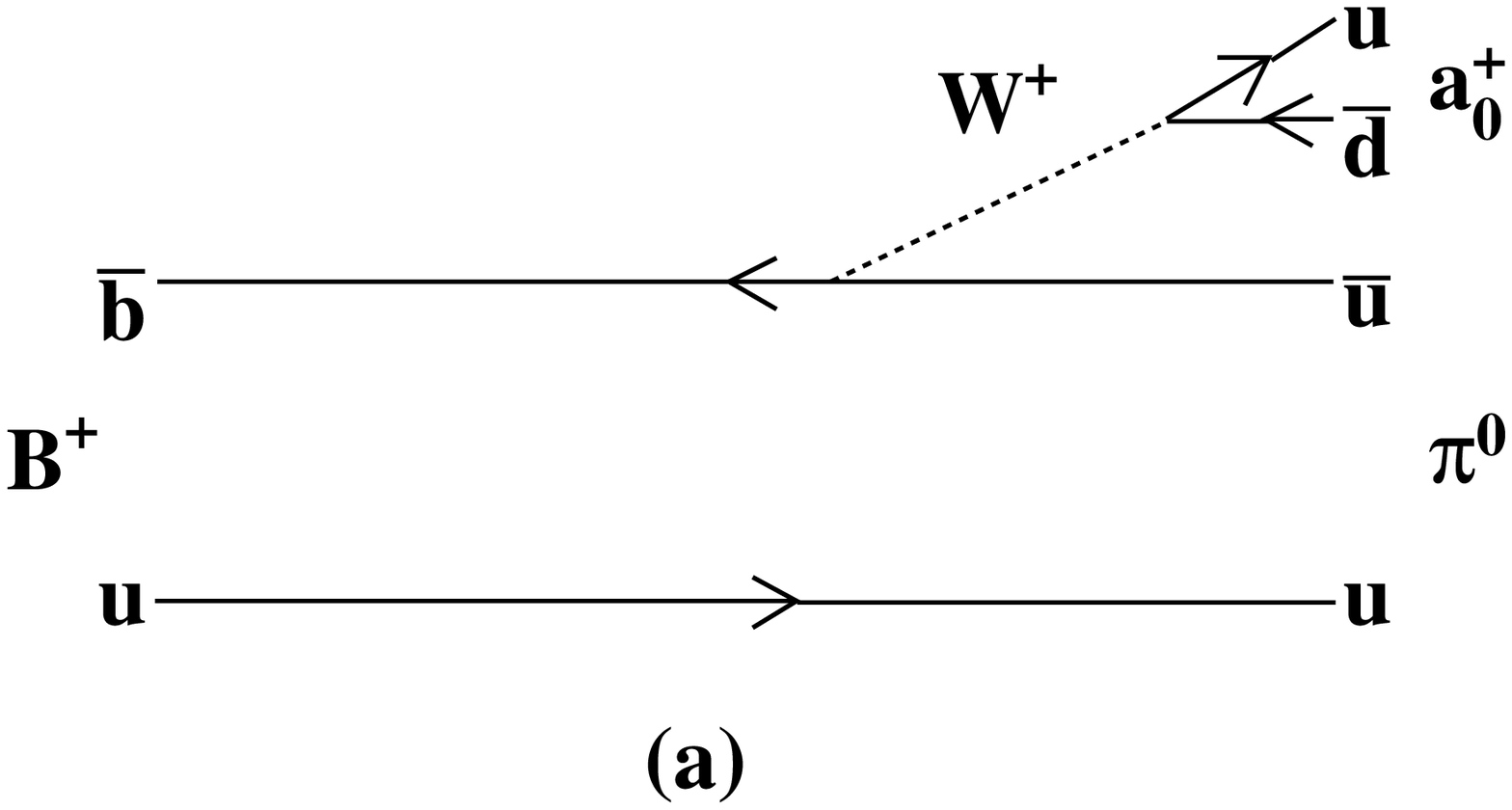}
\includegraphics[width=6.5cm]{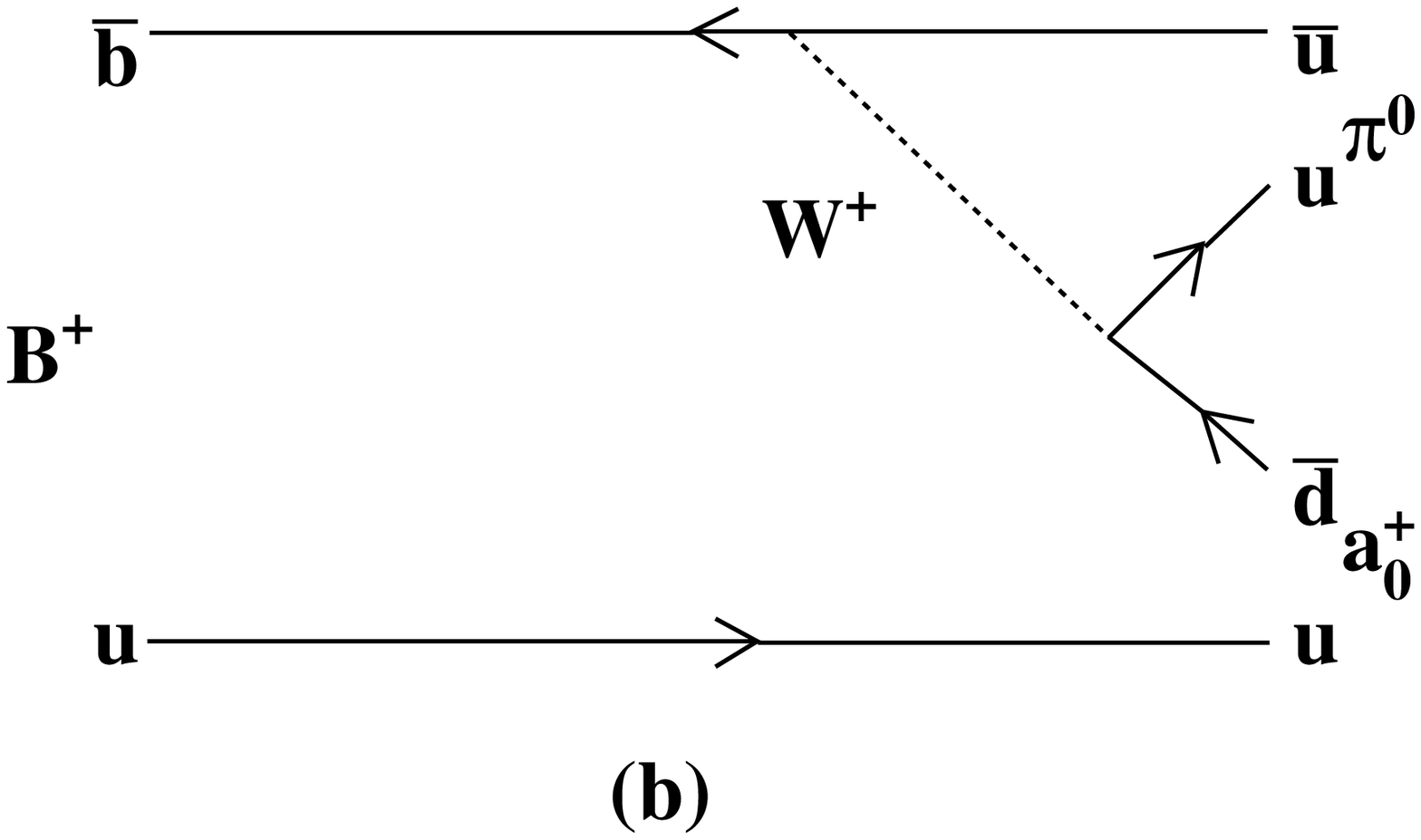}
\caption{\small{The proposed Feynman diagrams for the process \Btoazpiz\ with
(a) external (color allowed) production of the $\rm a_{0}^{+}$ and 
(b) internal (color suppressed) production of the $\rm a_{0}^{+}$.}}
\label{fig_feynmanDiagrams} 
\end{center} 
\end{figure}

Assuming the two quark model, the proposed Feynman diagrams for the
dominant tree level decays are given in
Fig.~\ref{fig_feynmanDiagrams}. The low expected branching fraction is
explained by the fact that the colour suppressed diagram (b) is
expected to dominate over the colour allowed diagram (a) since the
colour allowed case is also doubly suppressed by G-parity and vector
current conservation~\cite{ref:laplace2001}. Due to this it is also not possible to use
isospin arguments to relate the branching fraction for this mode to
others in the same final state Dalitz plane, such as $B^{+}
\rightarrow a_{0}^{0}\pip$.

The aim of the analysis described in this paper is to better constrain
the existing models by measuring the BF for the decay \Btoazpiz. The analysis
focuses solely on the case where the $a^{+}_{0}$ decays to an $\eta$ and a \pip,
with the $\eta$ decaying to two photons.  This sub-mode accounts for
$\approx 40\%$ of all $\eta$ decays.  An unbinned extended maximum
likelihood fit (ML) method is used to extract the $B^{+}\to
a^{+}_{0} \piz$ event yield.  Throughout this document the conjugate
decay mode $B^{-} \rightarrow a_{0}^{-}\piz$ is implied and $a_{0}$
refers to the $a_{0}(980)$ unless stated otherwise.

\section{THE \babar\ DETECTOR \& DATASET}
\label{sec:babar}

The analysis was performed using data collected with the \babar\ detector
\cite{ref:babar} at the PEP-II asymmetric-energy $e^{+}e^{-}$ collider
\cite{ref:pep} at the Stanford Linear Accelerator Center.  
Charged particles are detected and their momenta measured with a
5-layer double sided silicon vertex tracker (SVT) and a 40-layer drift
chamber (DCH) inside a $ 1.5 \rm{\, T}$ solenoidal magnet. A quartz
bar ring-imaging \Cerenkov\ detector (DIRC) complements the \dedx\
measurement in the DCH for the identification of charged
particles. Energies of neutral particles are measured by an
electromagnetic calorimeter (EMC) composed of $6,580$ CsI(Tl)
crystals, and the instrumented magnetic flux return (IFR) is used to
identify muons.

The data sample consists of 229.8\invfb collected at the $\Upsilon$(4S)
resonance. The number of \BB pairs used in the analysis totals
$(252.2 \pm 2.8)\times 10^{6}$~\cite{ref:bcounting}.

\section{RECONSTRUCTION \& SELECTION}
\label{sec:selection}
The analysis was based on the reconstruction of final states consisting of
four photons and one \pip. All charged particle tracks were taken as
\pip candidates unless they were identified as kaons by a particle
identification algorithm based on DCH and DIRC information. Tracks
were required to have momentum $\leq$ 10\gevc with
transverse momentum $\geq$ 0.1\gevc.  They were also required to be
composed from $\geq$ 12 DCH hits and with a distance of closest
approach to the interaction region within 1.5\cm in the transverse
direction and 10\cm along the axis of the detector.

In order to construct $\eta$ candidates, all pairs of photons in each
event were combined. The individual photons were required to have
energy between 0.05 and 10\gev. The invariant mass of each candidate
pair was then required to be between 0.515 and 0.569\gevcc. This
required mass range was optimised by maximising the ratio of signal to
the square root of the background, $S/\sqrt{B}$, using signal and
background Geant4 Monte Carlo (MC) samples.

The $\eta$ candidates were refitted to constrain their mass to the known 
value~\cite{ref:PDG2004} and were then
combined with the \pip candidates
to form $a^{+}_{0}$ candidates, which were
required to have a mass between 0.8 and 1.2\gevcc. The
$a^{+}_{0}$ mass selection was left relatively loose as this variable
was used in the ML fit.

Pairs of photons in the event were also used to form \piz
candidates. The photons were required to have energy between 0.03 and
10\gev.  The \piz candidates were required to be consistent with the
mass of a \piz, i.e. in the range 0.115 to 0.150\gevcc. The \piz
candidates were also refitted to constrain their mass to the known
value~\cite{ref:PDG2004}.

$B$ meson candidates were formed by combining all $a^{+}_{0}$ and \piz
candidates in the event. The $B$ candidates were described
kinematically using the energy substituted mass $\mes = [(\frac{1}{2}s
+ \vec p_{0}.\vec p_{B})^{2}/{E_{0}^{2}} - |\vec
p_{B}|^{2}]^{\frac{1}{2}}$ and the energy difference $\DeltaE = E_{B}^* -
\frac{1}{2}\sqrt{s}$. Here, $\vec p_{B}$ and $E_B$ are the momentum and energy of the $B$ candidate, $\vec p_{0}$ and $E_0$ are
the momentum and energy of the initial $(e^{+}, e^{-})$ state and the
$^*$ indicates this quantity is expressed in the centre-of-mass
system. $\sqrt{s}$ is the CM energy.  Both \mes and $\DeltaE$ were
included in the ML fit with the requirements, 5.20 $<$
\mes $<$ 5.29\gevcc and $|\DeltaE|$ $<$ 0.35\gev.  Since event reconstruction
is an imperfect process more then one candidate was produced for each
$B$.  The average number of $B$ candidates per signal event was
observed in signal MC to be 1.38. Further selection based on the
quality of these candidates was not performed and all were included in
the ML fit sample. The fit was designed, as explained and verified in
section 5, to only pick up one candidate per event.

\section{BACKGROUNDS}
\label{sec:backgrounds}
The major background in this analysis comes from random combinations
of particles in continuum $e^{+}e^{-}\rightarrow q\overline{q}$ events
($q$ = $u$,$d$,$s$,$c$). These were primarily rejected by placing a
requirement on the cosine of the angle between the thrust axis of the
reconstructed $B$ candidate and the thrust axis of the rest of the
event, $|\cos(\theta_{TB})|$; the thrust was calculated in the CM
frame. The distribution of this variable peaks sharply near 1.0 for
the two jet-like combinations from continuum $q\overline{q}$ pairs and is
approximately uniform for $B$ meson pairs; $|\cos(\theta_{TB})|$ was
required to satisfy $|\cos(\theta_{TB})| < 0.594$. This requirement
was chosen by again maximising $S/\sqrt{B}$ using signal and
background MC samples.

In order to further separate signal from continuum a Fisher
discriminant $\cal{F}$ \cite{ref:fisher} was used in the ML fit. The
discriminant was a weighted linear combination of the absolute value
of the cosine of the angle between the direction of the $B$ candidate
momentum and the beam axis, $|\cos(\theta_{TB})|$ as defined above,
and the $L_{0}$ and $L_{2}$ Legendre polynomial projections of the
energy flow of the event with respect to the $B$ candidate thrust
axis. The $L_{0}$ and $L_{2}$ quantities were formed by summing over
all the neutral and charged particles in the event which were not used
to form the $B$ candidate. The Fisher discriminant was required to
satisfy $-3 < {\cal F} < 1$.

The $B\overline{B}$ background was split into charmed and charmless
decay components, both of which were accounted for in the fit. The
charmed component was modelled by removing the charmless component
from the \BB MC sample. The charmless contributions were selected by
studying $B$ decays in MC with identical or similar final states to
the signal decay. Further contributions were identified from general
charmless $B$ decay MC samples which pass the selection
criteria. These modes were modelled using exclusive MC samples. From
the final state Dalitz plot we expect only 1 significant peaking
background ($\Bpm\to\rho^{\pm}(1450)\eta$) to overlap our signal.  The
other modes which contribute are mainly present through some degree of
mis-reconstruction. All B background yields were held fixed at their
expected values in the final fit. The expected contributions for these
modes are listed in Table~\ref{tab:charmless_bkgds}.

\begin{table} 
\begin{small} 
\caption{\small{The charmless modes identified as potentially
contributing significantly to the background. The nominal contributions
used in the fit are given.}}
\vspace{0.1cm} 
\begin{center}
\begin{tabular}{l|c} 
\hline   
Decay Mode & Expected Yield (B candidates)\\
\hline
\hline
Charm B decays & \\
\hline
Charged $b\rightarrow c$                     & $322 \pm 32$ \\
Neutral $b\rightarrow c$                     & $200 \pm 20$ \\
\hline
Charmless B decays & \\
\hline
$\Bpm \to\rho^{\pm}\piz$            &  $78 \pm 10 $\\
$\Bpm\to a^{\pm}_{1}\piz$           &  $58 \pm 13$\\
$\Bpm \to \rho^{\pm} \eta$          &  $37^{+8}_{-7} $\\
$\Bpm\to\rho^{\pm}(1450)\eta$       &  $25 \pm 25$\\
$\Bz \to \piz\piz$                  &  $19 \pm 4 $\\
Inclusive $\B \to X_{s}\gamma$      &  $18^{+5}_{-4} $\\
$\Bz \to \eta \pi^{0}$              &  $14 \pm 14 $\\
$\Bz \to a^{\pm}_{0} \rho^{\mp}$    &  $12 \pm 12$\\
$\Bpm \to a^{0}_{0} \rho^{\pm}$     &  $6 \pm 6$\\
$\Bpm \to a^{\pm}_{0} (1450) \piz$  &  $5 \pm 5$\\ 
$\Bpm \to \pi^{\pm} \eta \piz$ (non-resonant)&  $2 \pm 2$\\
$\Bpm\to\pi^{\pm}\piz\piz$ (non-resonant)    &  $2 \pm 2$\\
\hline 
\end{tabular}
\label{tab:charmless_bkgds}  
\end{center}
\end{small} 
\end{table}

\section{MAXIMUM LIKELIHOOD FIT}
\label{sec:thefit}
The input variables to the extended ML fit were
\mes, $\DeltaE$, $\cal{F}$, and the $a^{+}_{0}$ resonance mass. The
probability density functions (PDF) used to model the data are
discussed below, and were determined from MC simulation. The exception
to this is the continuum component where most parameters were left
free in the fit, thus enabling us to extract the parameters directly
from the data. The only other floated quantities in the fit
were the signal and continuum yields.

\subsection{Signal Model}
\label{sec:Signal Model}
The peaking components in $\DeltaE$ and \mes were modelled with a
Novosibirsk function~\cite{ref:novosibirsk} and two independent
Gaussians respectively. The $\cal{F}$ shape was modelled using an
asymmetric Gaussian and the $a^{+}_{0}$ resonance mass with a
Breit-Wigner. The resonance was modelled in signal MC with a mass of
0.98\gevcc and a width of 80\mevcc~\cite{ref:colorado}. 

The signal shape can be distorted by true signal candidates which have been
mis-reconstructed and is referred to as self-crossfeed (SXF). The $B$
candidates affected have more background-like distributions in the fit
variables. If improperly accounted for these can result in a reduction
of the discriminating power of the fit.

\begin{figure}[htbp] 
\begin{center}
\includegraphics[width=12cm]{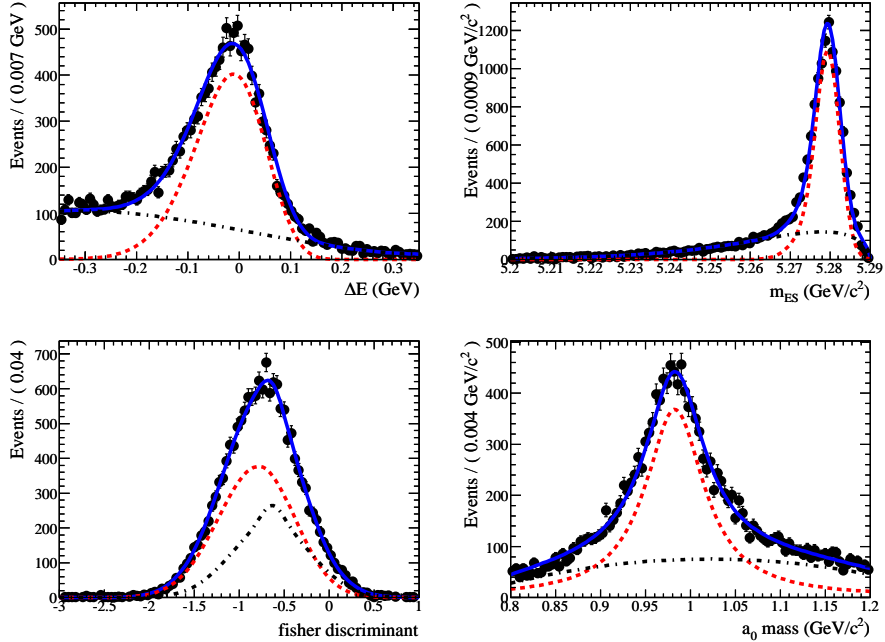}
\caption{\small{Signal model PDFs projected onto signal MC. The total PDF is given in blue (solid line) with the true signal contribution in red (dashed) and the SXF in black (dash-dotted). Only the true signal shape is used in the final fit.}}
\label{fig_signalmodel} 
\end{center} 
\end{figure}

In some signal events one true $B$ candidate was seen as well as a
number of SXF $B$ candidates. As stated in Section 3 this gives an
average of 1.38 $B$ candidates per event.  In order to remove any
distortion which may result from SXF, therefore giving the purest
possible signal shape, a model was created which combines the expected
true signal shape (described above) with a model to describe the
SXF. High purity samples of signal and SXF were used for this purpose.
The true signal sample was obtained from MC by requiring that all
generated daughter particles of the signal decay were correctly
reconstructed.  This method was not 100$\%$ efficient, leading to
cross-contamination between the signal and SXF samples. In order to
minimise this effect the signal and SXF samples were fitted
iteratively to determine the PDF parameters.  The SXF model was not
explicitly included in the final fit since its purpose was to
facilitate the removal of SXF distortion from the signal shape. The
resulting PDFs, projected onto signal MC, are presented in
figure~\ref{fig_signalmodel}.

It was verified that there was negligible contamination of the signal
component with SXF events and that these were instead absorbed by the
continuum PDF.  It was also verified that this method yields no
greater than one good signal candidate per event, despite the
inclusion of multiple candidates.  The tests are outlined in
Section~\ref{sec:Fit Validation}.

\subsection{Background Model}
\label{sec:Background Model}
Any peaking contributions in \mes and $\DeltaE$ in the background
samples were modelled using Gaussian shapes. Where a slowly varying,
combinatorial component exists it was modelled with low-order
polynomials in $\DeltaE$ and an ARGUS \cite{ref:argus} threshold
function for \mes. In most charmless modes correlations were expected
to exist between these variables. In order to model these correlations
the above prescription was, where necessary, replaced with a
two-dimensional non-parametric KEYS PDF \cite{ref:keys}. $\cal{F}$ was
modelled either with two Gaussians or an asymmetric Gaussian. Peaking
components in the $a^{+}_{0}$ mass variable were modelled with
Gaussians or Breit-Wigners depending on mis-reconstruction. Any more
slowly varying components were modelled with low-order polynomials.

\subsection{Fit Validation}
\label{sec:Fit Validation}
In order to verify the consistency of the fitting procedure MC
simulations were generated from the PDF model. By refitting to these
datasets and assessing the shifts in the fitted signal yields any
problems with the model could be detected. In the studies no
significant deviation from the generated signal yield was found. In order to
detect biases from any improperly modelled correlations among the
variables the above study was repeated with embedded randomly selected
events from the fully simulated signal and background MC samples. No
significant biases from any of the background modes were found, nor
from any correlations in signal MC.  Any small biases which exist were
used to estimate a systematic error contribution. Finally, in order to
test the ability of the fit to separate true signal from SXF, SXF
candidates were explicitly embedded from the signal MC and a check for
shifts in the fitted signal yield was made. Once again the fitted signal
yields were consistent with those expected.

\section{SYSTEMATIC STUDIES}
\label{sec:Systematics}
The systematic uncertainties fall into three categories: firstly the
errors in the number of events extracted from the fit, secondly the errors
in the efficiency, and lastly the errors in the BF result due to
uncertainties in the total number of B mesons 
and in the daughter BFs. 

\subsection{Systematic errors derived from the fit}
Uncertainties resulting from the ML fit give additive systematic errors
in the number of signal events derived from the fit.

\begin{itemize}
\item{\underline{Uncertainties in the PDF parameters used in the fit}\\
This uncertainty lies in the statistical errors in the fitted
parameters of the PDFs which make up the model. These were due to the
limited amount of available MC simulated events upon which to model
any given PDF.  Each parameter in a PDF was considered independent of
the others, thus neglecting any correlations which may exist.  The
PDFs considered were signal, \BB events with charmless modes removed,
and continuum. The charmless contribution was expected to be small and
was neglected.  With the exception of the signal $a_{0}$ mass
Breit-Wigner width, all of the parameters were varied within their
errors and the final fit repeated.  In all cases the resulting shifts
in the fitted signal yield were taken as the systematic error
contribution. For the signal $a_{0}$ mass Breit-Wigner width lower and
upper bounds of 50\mevcc and 100\mevcc were assumed. This was to account
for the uncertainty in the modelled $a_{0}$ lineshape.}

\item{\underline{Uncertainties in charmless $B$ decay contributions}\\
The uncertainty here lies in the normalisation of the separate
charmless backgrounds due to errors in the BFs used to estimate
their contribution. 
These values are listed in Table~\ref{tab:charmless_bkgds}.
The systematic error contribution was calculated as the shift in the
fitted signal yield in the final fit to data where the contributions
from each specific mode were varied within the errors of their
BFs. For the cases where only an upper limit exists, the contributions
were calculated assuming 50\% of the limit as a central value. 
The
systematic error was then estimated by varying this assumed central value by
$\pm 100\%$.}

\item{\underline{Uncertainties in the yields of the $B\to$ charm modes}\\
This uncertainty was ascertained by varying the yields of the
$B$ to charmed modes by $\pm10\%$ as a conservative estimate.
This yield uncertainty was
dominated by the error in the efficiency for reconstructing these $B$
modes
and not the cross-section for their production. 
Once again the shifts in the signal yield
were taken as systematic errors.}

\item{\underline{Uncertainties in the bias from the fit}\\ 
To estimate any potential fit bias,
a simulation study was run embedding all $B$
backgrounds from the MC and generating continuum MC according to the
values of the continuum shape parameters resulting from the final
fit. Zero signal events were embedded or generated. The
systematic error was estimated as the sum in quadrature of 50$\%$ of the
fitted bias and its statistical error.}
\end{itemize}

\subsection{Systematic errors in the efficiency}
There are a number of sources of systematic uncertainty which affect
the efficiency and are thus applied as a multiplicative correction to
the final result.
\begin{itemize}
\item{\underline{Tracking and neutrals efficiencies}\\
This source of uncertainty arises in tracking where a global per track
systematic error of $0.5\%$
was assigned based on results of dedicated studies.
There was also an uncertainty in the efficiency for
reconstructing neutral particles. A systematic error of
$3\%$ each for every \piz and $\eta$ in the final state was assigned.}

\item{\underline{Data/MC agreement in the $|\cos{\theta_{TB}}|$ variable}\\ 
Due to the imperfect agreement between data and MC samples, the
selection criteria for $|\cos{\theta_{TB}}|$ require the assignment of
a systematic error. Control sample studies have shown that tighter
selections incur larger errors. As such a conservative 5$\%$ error due
to this selection was assigned.}

\item{\underline{Statistical error}\\
Finally, the error due to limited MC statistics in the
efficiency had to be accounted for. This was simply the binomial error
in selecting a given number of events from a larger sample.}
\end{itemize}

\subsection{Systematic errors contributing to the branching fraction}
These systematic contributions are also multiplicative errors.
\begin{itemize}
\item{\underline{Total number of \BB events}\\
The total number of \BB events in the data set was estimated to be
$(252.2 \pm 2.8)\times10^{6}$. The error was taken as a systematic.} 

\item{\underline{Uncertainties in the daughter decay BFs}\\
The errors in ${\BR(a_{0}^{+} \to \eta\pip)}$ and ${\BR(\eta
\to \gamma \gamma)}$ were taken from the Particle Data Group~\cite{ref:PDG2004}. 
The error in ${\BR(a_{0}^{+} \to \eta\pip)}$ was calculated by taking
the ratio of the partial widths of the two dominant $a_{0}^{+}$ decay
modes; $a_{0}^{+} \to \eta\pip$ and $a_{0}^{+} \to K\overline{K}$. The
measured ratio is $0.183\pm0.024$ which, assuming all other decay
modes are negligible, gives ${\BR(a_{0}^{+} \to \eta\pip)} = 0.85 \pm
0.02$.  The value taken for ${\BR(\eta \to \gamma \gamma)}$ was
$0.3943\pm 0.0026$.}
\end{itemize}

\subsection{Summary of Systematics}

The results of all of the studies to estimate systematic error are
presented in Table~\ref{tab:systematics}.

\begin{table}[!htbp]
\caption{\small{Estimated systematic errors in the final fit result.}}
\vspace{0.2cm}
\small
{\centering \begin{tabular}{l|c}
\hline
Source of Uncertainty      & $\eta \to \gamma\gamma$      \\
\hline\hline
Additive (Events)                     &                   \\
\hline
Fit Parameters                        & $^{+7.7}_{-4.4}$  \\
Charmless Yields                      & $^{+2.3}_{-1.5}$  \\
Charm Yields                          & $^{+0.2}_{-0.0}$        \\
Fit Bias	                      & $\pm 1.6$         \\
\hline
Total Additive (Events)               & $^{+8.2}_{-4.9}$  \\
\hline\hline
Multiplicative ($\%$)                 &                   \\
\hline
Neutral efficiency                    & $\pm 6.0$         \\
Tracking efficiency                   & $\pm 0.5$         \\
$|\cos(\theta_{TB})|$ Selection        & $\pm 5.0$         \\
MC Statistics                         & $\pm 0.9$         \\
Number of $B\overline{B}$ Events      & $\pm 1.1$         \\
Daughter $a_{0}$ Decay BF             & $\pm 2.0$         \\
Daughter $\eta$ Decay BF              & $\pm 0.7$         \\
\hline
Total Multiplicative ($\%$)           & $\pm 8.2$         \\
\hline
\end{tabular}\par}
\label{tab:systematics}
\end{table}

\section{RESULTS}
\label{sec:results}
The results from the fit to the full data set are given in
Table~\ref{tab:results}. The yield is found to be consistent with
zero. 

\begin{table}[h!] 
\begin{small} 
\caption{\small{The results of the fit to the full data set and other values required 
for calculating the branching fraction. All B background yields were
held fixed to the values listed in Table 1. The upper limit is shown
first with only the statistical error and then with the total error.}}
\vspace{0.1cm} 
\begin{center}
\begin{tabular}{l|c} 
\hline
Required quantity/result    &      \\
\hline\hline
Candidates to fit          & 36098       \\
Signal Yield (events)      & -18$\pm$11  \\
Continuum Yield (candidates)   & 35324$\pm$190\\
ML Fit bias (events)       & 2.55        \\
\hline
Accepted eff. and BFs      & \\
\hline
$\epsilon$ ($\%$)	   & 16.18       \\
${\cal{B}}(\eta\to \gamma\gamma)$ ($\%$)   & 39.43  \\
${\cal{B}}(a^{+}_{0}\to\eta\pip)$ ($\%$) & 84.5   \\
\hline
Results                    & \\
\hline
Branching Fraction  ($\times10^{-6}$) & $-1.5^{+0.9}_{-0.7} (stat) ^{+0.6}_{-0.4} (syst)$\\
Upper Limit 90$\%$ C.L. ($\times10^{-6}$) & $<$ 1.06 (statistical error only)\\
Upper Limit 90$\%$ C.L. ($\times10^{-6}$) & $<$ 1.32 (total error)\\
\hline
\end{tabular}
\label{tab:results}  
\end{center}
\end{small} 
\end{table}

\begin{figure}[!h] 
\begin{center}
\includegraphics[width=7cm]{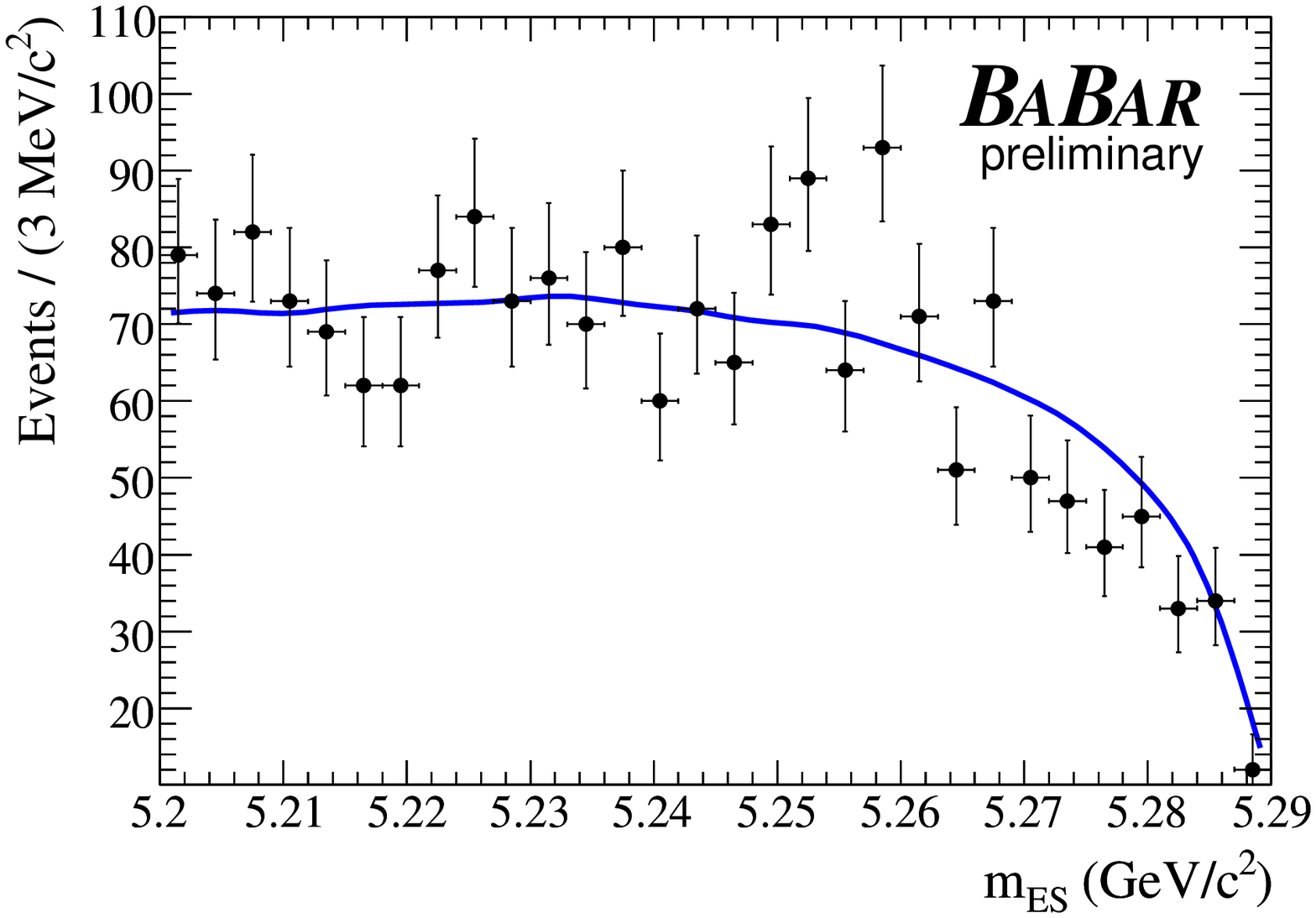}
\includegraphics[width=7cm]{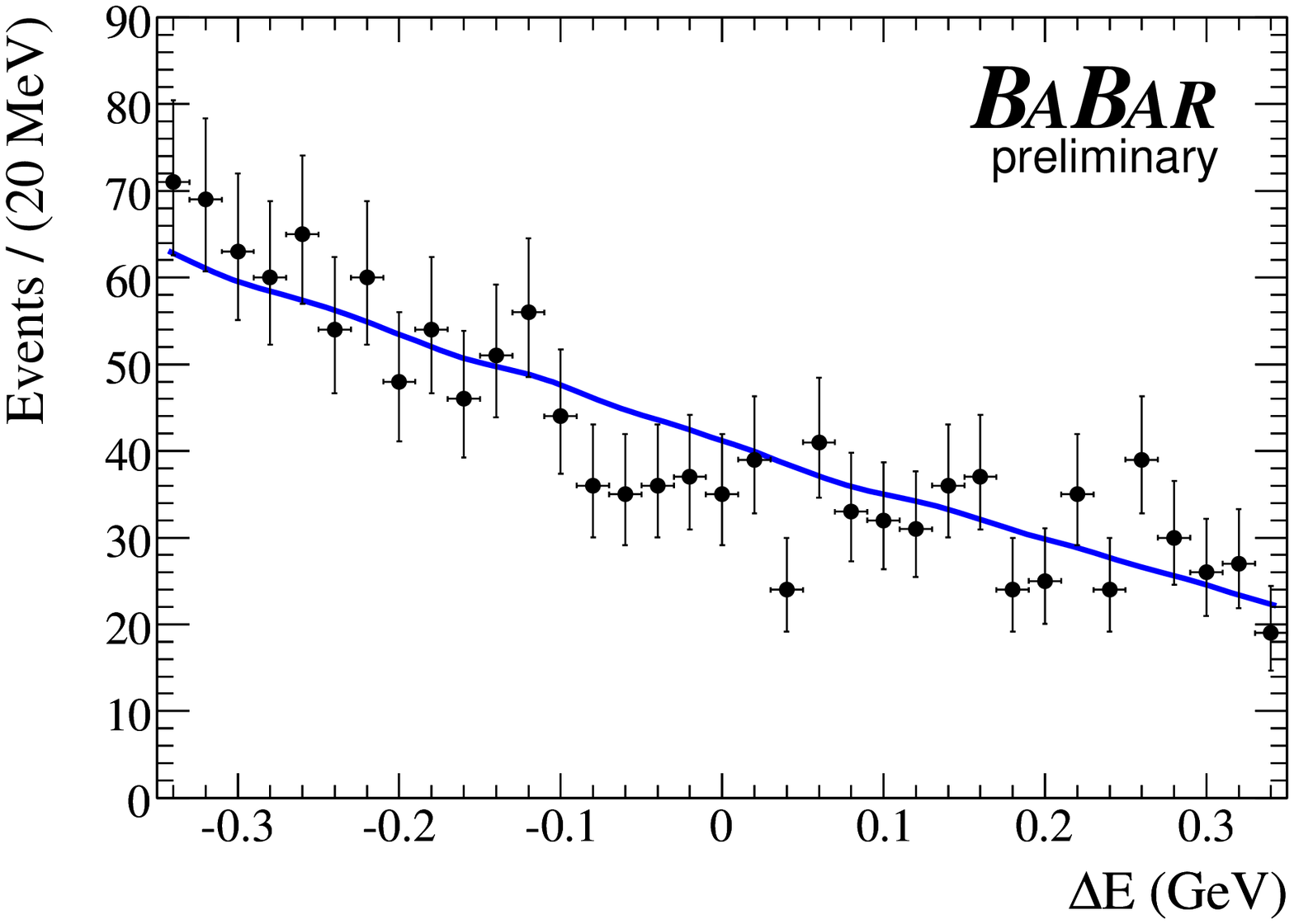}
\caption{\small{Projection plots in \mes and $\DeltaE$ comparing MC generated from the overall PDF (blue solid curve) with the data (black points). The samples have been background reduced by requiring the ratio of likelihoods ${\cal{L}}_{sig}/[{\cal{L}}_{sig} + \Sigma{\cal{L}}_{bkg}]$ to be $>$ 0.6.}}
\label{fig_projections} 
\end{center} 
\end{figure}

\begin{figure}[!h] 
\begin{center}
\includegraphics[width=12cm]{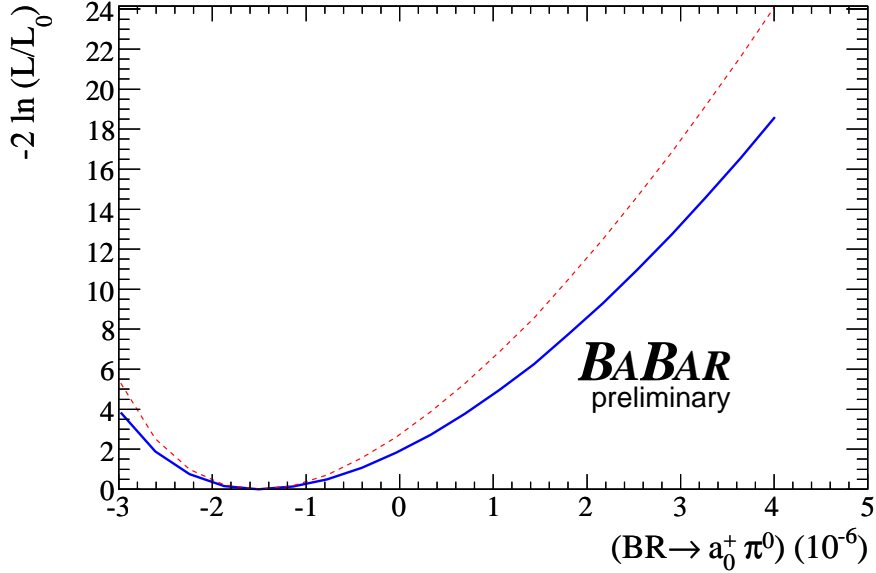}
\caption{\small{Log likelihood scan curve showing the effect of convoluting 
in systematic errors. The red dashed curve shows the likelihood for statistical
errors only while the blue solid curve shows the likelihood including
systematic errors.}}
\label{fig_Lscan_gg} 
\end{center} 
\end{figure}

The branching fraction is
calculated using the following formula
\begin{eqnarray} 
{\cal{B}} = \frac{Y - Y_{B}}{\epsilon  N_{B\overline{B}} \, {\Pi}_i {\cal{B}}_{i} }
\end{eqnarray}
\noindent
where $Y$ is the signal event yield from the fit, $Y_{B}$ is the fit
bias, $\epsilon$ is the efficiency for the $B$ decaying via the studied
mode, $N_{B\overline{B}}$ is the number of produced $B\overline{B}$
mesons and $\Pi_i {\cal{B}}_i$ is the product of the daughter
branching fractions. The decay rates for $\Upsilon(4S)\to\BpBm$ and $\Upsilon(4S)\to\BzBzb$ are assumed to be equal.

Projection plots for the final fit results are presented in
Figure~\ref{fig_projections}. These compare a sample of MC generated
from the overall PDF with the data. Both samples have had background
reduced by requiring that the likelihood ratio
${\cal{L}}_{sig}/[{\cal{L}}_{sig} + \Sigma{\cal{L}}_{bkg}]$ for any event be $>$ 0.6.

The likelihood $\cal{L}$ was convoluted with a Gaussian to include the
systematic errors. A scan was performed to give the variation in the
likelihood for different input yields. The resulting curve in
$-2\ln({\cal{L}}/{{\cal L}_{0}})$ as a function of the branching fraction is
shown in Fig.~\ref{fig_Lscan_gg}. Note how this plot is scaled
relative to the value for the minimum negative log likelihood
(${\cal{L}}_{0}$).  From this distribution the upper limit of $1.32
\times 10^{-6}$ was computed by integrating the associated likelihood
function to find the yield which bounds 90$\%$ of the area under the
curve. In doing this a starting branching fraction value of zero was
assumed.

This upper limit is reported in Table~\ref{tab:results} with and
without the systematic error to indicate the degree to which the limit
is statistics dominated. The largest source of systematic error comes 
from the $a_{0}$ lineshape.

\section{SUMMARY}
\label{sec:summary}
A search for the decay \Btoazpiz was carried out using 252 million
\BB pairs. An unbinned extended maximum likelihood fit was used to
obtain the result. The fitted signal yield was consistent with zero
and thus a 90$\%$ C.L. upper limit of $1.32 \times10^{-6}$ was set on
the branching fraction for this decay. Our sensitivity to this mode
was therefore insufficient to make it possible to differentiate
between the two and four quark $a_{0}$ structure models as defined
by Delepine {\em et al.}~\cite{ref:delepine2005}.

\section{ACKNOWLEDGMENTS}
\label{sec:acknowledgments}
The authors are grateful for the 
extraordinary contributions of our \pep2\ colleagues in
achieving the excellent luminosity and machine conditions
that have made this work possible.
The success of this project also relies critically on the 
expertise and dedication of the computing organizations that 
support \babar.
The collaborating institutions wish to thank 
SLAC for its support and the kind hospitality extended to them. 
This work is supported by the
US Department of Energy
and National Science Foundation, the
Natural Sciences and Engineering Research Council (Canada),
Institute of High Energy Physics (China), the
Commissariat \`a l'Energie Atomique and
Institut National de Physique Nucl\'eaire et de Physique des Particules
(France), the
Bundesministerium f\"ur Bildung und Forschung and
Deutsche Forschungsgemeinschaft
(Germany), the
Istituto Nazionale di Fisica Nucleare (Italy),
the Foundation for Fundamental Research on Matter (The Netherlands),
the Research Council of Norway, the
Ministry of Science and Technology of the Russian Federation, and the
Particle Physics and Astronomy Research Council (United Kingdom). 
Individuals have received support from 
the Marie-Curie IEF program (European Union) and
the A. P. Sloan Foundation.


\begin{thebibliography}{99}
\bibitem{ref:novosibirsk}
The Novosibirsk function is defined as $f(x)=A_{s} \exp(-0.5(\ln^{2}[1+ \Lambda\tau(x - x_{0})]/\tau^{2} + \tau))$
where $\Lambda = \sinh(\tau\sqrt{\ln 4})/(\sigma\tau\sqrt{\ln 4})$, the peak is $x_{0}$ and $\tau$ is the tail parameter.

6~
\bibitem{ref:argus}
The ARGUS function is defined as $f(x) = x{[(1-(x/E_{beam}))^{2}\exp{[p(1-(x/E_{beam}))^{2}]}]}^{1/2}$.
\end{thebibliography}
\end{document}